\def\minone{$^{-1}$}
\def\sqiggt{\hbox{\rlap{\lower.55ex \hbox {$\sim$}}
\kern-.3em \raise.4ex \hbox{$>$}\,}}
\def\sqiglt{\hbox{\rlap{\lower.55ex \hbox {$\sim$}}
\kern-.3em \raise.4ex \hbox{$<$}\,}} 
\def\kev{\,ke\kern-.1em V} \def\ev{\,e\kern-.1em V} 
\def\sqig{$\sim\,$} \def\etal{et\,al.} \def\msun{M$_{\scriptstyle\odot}$} 
\def\up#1{$^{\mbox{{\scriptsize #1}}}$}  
 \def\pten#1{$\times10^{#1}$}
\def\deg{$^{\circ}$}  

\def\xte{{\sl RXTE\/}}\def\beat{$\omega$\,$-$\,$\Omega$}
\def\ex{EX~Hya}
\def\ginga{{\it Ginga\/}}
\documentstyle{mn}
\title[Outbursts of EX~Hydrae]
{Outbursts of EX~Hydrae: mass transfer events or disc instabilities?}
\author[C.~Hellier \etal]{Coel Hellier,\up{1}\ Jonathan Kemp,\up{2}\
T. Naylor,\up{1}\ Frank M. Bateson,\up{3}\ Albert Jones,\up{3}\cr
Danie Overbeek,\up{3}\ Rod Stubbings\up{3}\ and Koji Mukai\up{4}\\
\up{1}Department of Physics, Keele University, Keele, Staffordshire, ST5 5BG\\
\up{2}Astronomy Programs, Biosphere 2 Center, Columbia University, 32540 S Biosphere Rd,
Oracle, AZ 85623, U.S.A.\\
\up{3}Variable Star Section, Royal Astronomical Society of New Zealand,
PO Box 3093, Greerton, Tauranga, New Zealand.\\
\up{4}Laboratory for High Energy Astrophysics, NASA/Goddard Space Flight
Center, Greenbelt, MD 20771, U.S.A.}

\date{ }
\begin{document}
\maketitle
\begin{abstract}
We present the 45-yr record of \ex 's lightcurve and discuss the 
characteristics of its 15 observed outbursts. We then concentrate on the
1998 outburst, reporting the first outburst X-ray observations. We discover 
an X-ray beat-cycle modulation, indicating that an enhanced accretion stream
couples directly with the magnetosphere in outburst, confirming our previous 
prediction. Optical eclipse profiles late in outburst show that the 
visible light is dominated by an enhanced mass-transfer stream overflowing
the accretion disc. We are uncertain whether the enhanced mass transfer 
is triggered by a disc instability, or by some other cause. 
While in outburst, \ex\ shows some of the characteristics of SW~Sex stars.
\end{abstract}

 
\section{Introduction}
If intermediate polars (IPs) are cataclysmic variables possessing partial
accretion discs, with the centre disrupted by a magnetic field, then
we expect that they can show disc instability outbursts, as dwarf novae
do. Several IPs --- XY~Ari (Hellier, Mukai \&\ Beardmore 1997), 
YY Dra (Patterson \etal\ 1992) and GK~Per (e.g.\ Kim \etal\ 1992) 
--- appear to show just that. 

However, two other IPs --- V1223~Sgr \&\ TV~Col --- show
short, low-amplitude outbursts that are unlike dwarf nova eruptions
and probably result from another instability such as mass-transfer 
bursts (see Warner 1996 and Hellier \etal\ 1997 for reviews). 
The outbursts of TV~Col last \sqig 8 h with 2-mag amplitudes (Szkody 
\&\ Mateo 1984; Schwarz \etal\ 1988; Hellier \&\ Buckley 1993);
increased S-wave emission during this period points to enhanced mass 
transfer.
V1223~Sgr has shown a very similar event (van Amerongen \&\ van Paradijs 1989).
Since these two stars have novalike discs (e.g.\ TV Col shows 
superhumps and V1223~Sgr shows VY~Scl low states) the outbursts are 
unlikely to be thermal instabilities in the disc. Very similar short-lived
flare events have been seen in AM~Her stars (e.g.\ Warren \etal\ 1993),
which cannot be disc instabilities since such stars don't have discs.
[The possible IP RX\,J0757+6306 may be a similar system (Tovmassian \etal\ 1998;
Kato 1999), but the data are currently too sparse for certainty.]

The remaining IP showing outbursts, EX~Hya, is intermediate between the 
above two types, with outbursts lasting \sqig 2--3 d (cf.\ 0.5 d for 
TV~Col \&\ V1223~Sgr, and 5 d for XY~Ari \&\ YY~Dra). Previous outbursts have
been reported by Bond \etal\ (1987); Hellier \etal\ (1989; hereafter 
Paper~1); Reinsch \&\ Beuermann (1990) and Buckley \&\ Schwarzenberg-Czerny 
(1993).  

\begin{figure*} \vspace{18.5cm}      
\includegraphics{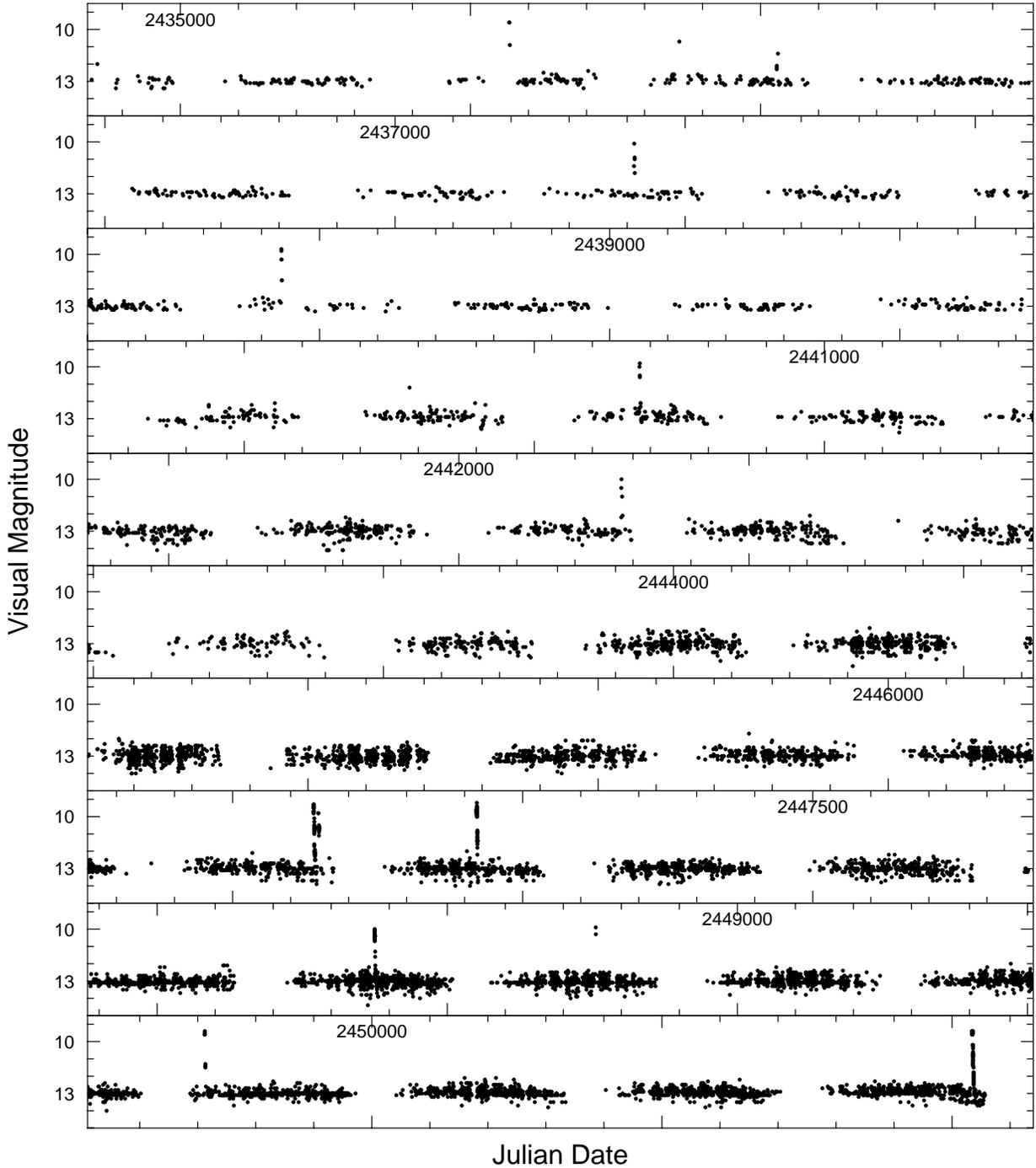}
\caption{The record of EX~Hya by the Variable Star Section of the Royal 
Astronomical Society of New Zealand.}
\end{figure*}

A major finding from the 1987 outburst (Paper~1)
was a high-velocity feature in the emission-line wings. This seemed to
arise from an enhanced accretion stream overflowing the disc and 
connecting directly with the magnetosphere. We predicted that since the 
stream rotated with the orbital frequency ($\Omega$) and the magnetosphere
with the spin frequency ($\omega$), the relative geometry (and thus
the X-ray emission from stream-fed accretion) should vary at the \beat\
frequency. Such X-ray beat modulations have since been seen in many 
IPs (e.g.\ Hellier 1998) but not, so far, in EX~Hya. Accordingly, we
applied for Target of Opportunity time to observe the next outburst
of EX~Hya with the rapid-response \xte\ X-ray satellite. This was 
successfully triggered during an outburst in 1998 August and we report
the results here. 

EX~Hya is also well studied in quiescence, showing a prominent sinusoidal
modulation at the 67-min spin period and a grazing eclipse recurring with the
98-min orbital period. See, e.g., Hellier (1987, hereafter Paper~2) for 
spectroscopy, Siegel \etal\ (1989) for optical photometry, and Rosen \etal\
(1991) for X-ray data.

\section{The long-term record}
Since \ex's behaviour is clearly different from that of normal dwarf novae
it is worth presenting the complete record. Fig.~1 shows the visual 
estimates of \ex\ compiled by the Variable Star Section of the RASNZ.    
While \ex\ mostly sits at 13\up{th}\ mag, it rises to mag 9.5 in infrequent 
outbursts lasting 2--3 d. Note that much of the variability at quiescence
is real, caused by the spin and orbital modulations. Fig.~2 contains 
details of the outbursts on an expanded scale. We note the following points:

\begin{enumerate}

\item 15 outbursts have been seen in 44 years, for an average recurrence
of $\approx$\,3 yrs. However, the yearly and monthly data gaps mean that 
many will have been missed. Coverage dense enough to catch 2-d outbursts
is $\approx$\,2/3\up{rds}\ complete in recent times, dropping to 
$\approx$\,1/3\up{rd}\ complete earlier, so we can estimate that only
$\approx$\,half the outbursts have been caught, reducing the recurrence to 
$\approx$\,1.5 yrs. 

\item The outbursts occur irregularly: near JD 244\,8370 a `double'
outburst occurred with an interval of only 8 d. In contrast, 
taking the sampling into consideration, there is a 95\%\ probability 
that inter-outburst intervals $>$\,2.7 yrs have occurred (most 
likely in the 12-yr period between JD 244\,2300--244\,6600 when no 
outburst was seen).

\begin{figure} \vspace{16.5cm}      
\includegraphics{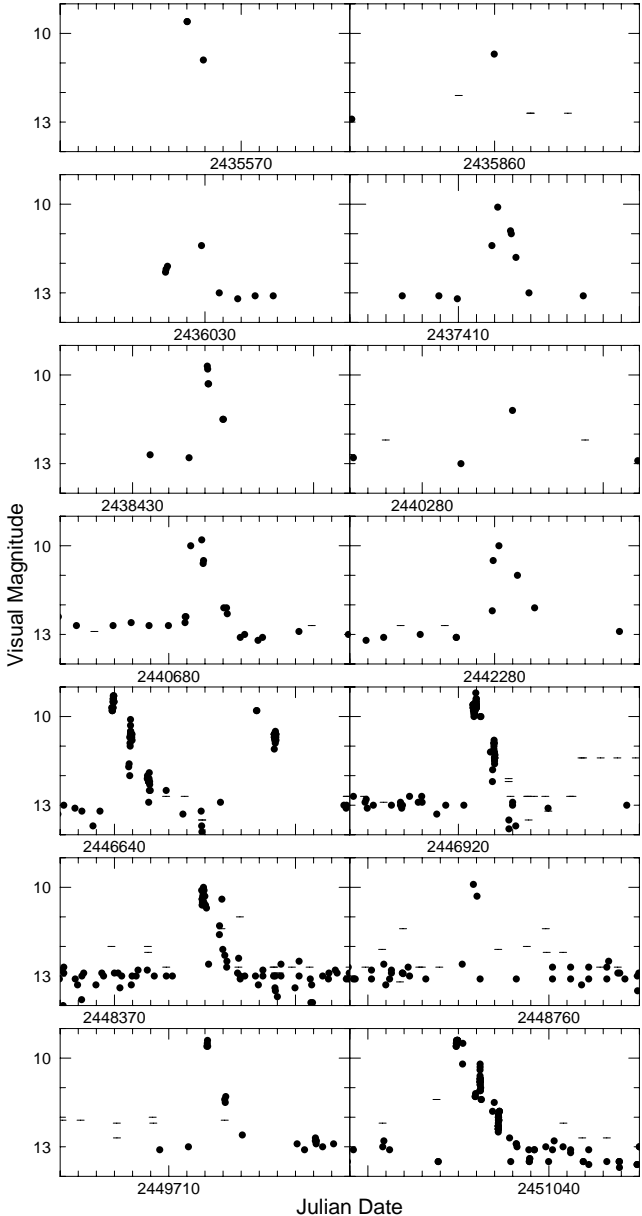}
\caption{Expanded plots of the outburst from Fig.~1. The tickmarks are
at 1-d intervals. Bars are upper-limits.}
\end{figure}

\item The outburst rises are unresolved in the RASNZ data, where rises of 
3.4 mags in \sqiglt 12 hrs are seen.  Reinsch \&\ Beuermann (1990) caught  
part of a rise, seeing the brightness increase by a factor 10
within 3 hrs. 

\item The declines are slower than the rises and are variable:
the outburst at JD 244\,6920 declined by 3.5 mags in 1.8 d
while that at JD 245\,1040 took 3.0 d to decline by the same
amount.

\begin{figure*} \vspace{9.5cm}      
\includegraphics{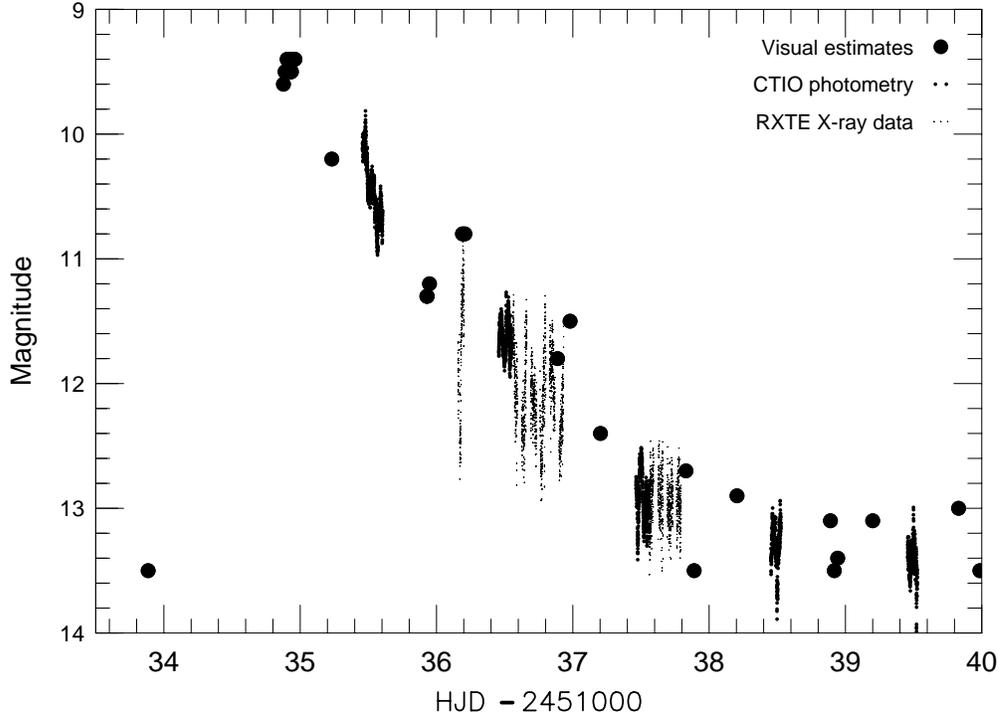}
\caption{The 1998 outburst of EX~Hya showing RASNZ visual estimates,
CTIO optical photometry and \xte\ X-ray data. The CTIO and \xte\ datasets
have had the zero-points adjusted to match the visual records.}
\end{figure*}

\item The outburst at JD 244\,8760 was peculiar. \ex\ rose from
mag 13.1 to 9.9 in $<$\,15 hrs and declined from 10.3 to 12.6
in only $<$\,4.5 hrs, the whole event being over in $<$\,1 d. 
[The two outburst points are by different observers; the
observers involved (including the current authors AJ and DO)
are highly experienced observers of EX~Hya, and 
the data points have been confirmed from the original observing logs.]

\item If the accretion rate scales as the optical magnitude and if
the quiescent accretion rate is \sqig 10$^{16}$ g\,s\up{--1}\ then the
outbursts typically involve \sqig 10$^{22}$ g of material. 

\end{enumerate}

\section{The 1998 outburst data}
The August 1998 outburst (Fig.~3) showed the usual unresolved rise but 
decayed to quiescence in 3 d, 1 d longer than the other well-studied outbursts.

\subsection{\xte\ X-ray observations}
Following notification of the outburst we observed \ex\ with 
\xte\ [see Bradt, Rothschild \&\ Swank (1993) for a description of this 
satellite] gaining three sections of data on the outburst decline
(Fig.~3). The first section, lasting 1 hr, recorded a 2--15 \kev\ count rate
varying between 70 and 330 c\,s\minone\ (all 5 PCUs); during the second
section, lasting 9 hr, the count rate was in the range 60--220; and by the
third section, lasting 6 hr, the count rate had declined to 
35--70, essentially a quiescent count rate. 

We Fourier transformed the 2--15 \kev\ X-ray dataset, first normalizing the
three data sections to the same count rate. The result (Fig.~4) reveals 
power at the spin frequency ($\omega$) and at the beat frequency between
the orbital and spin periods (\beat).  An X-ray beat frequency has never
been seen in \ex\ in quiescence, but its occurrence in outburst confirms 
the prediction in Paper~1. However, some scepticism is in
order since the X-ray data cover only \sqig 4.5 cycles of the 3.5-hr beat
period. One might also be concerned that since the spacecraft orbital
period is near EX~Hya's orbital period (96 vs 98 mins), beating with the 
spacecraft orbit might explain the peak seen. However, this would produce 
equal peaks at $\omega$\,$-$\,$\Omega_{\rm rxte}$ and 
$\omega$\,+\,$\Omega_{\rm rxte}$ whereas there is no power at 
$\omega$\,+\,$\Omega_{\rm rxte}$. 

By fitting a sinusoid to the three sections of 2--15 \kev\ X-ray data we 
find that the spin pulse had a modulation depth (semi-amplitude/mean) of
52\%\ in the first section, declining to 25\%\ in the second section and 5\%\
in the third (the errors are dominated by flickering, and the first result
is particularly unreliable since the data cover only \sqig 1 cycle). 
For comparison, Rosen \etal\ (1991) quote a 14\%\ depth in quiescent 
\ginga\ data over a similar energy range; thus the pulse amplitude was 
markedly bigger during outburst.  There was no apparent change in pulse
phase during the \xte\ observations.

The spectral changes over the spin cycle are consistent with the
usual quiescent behaviour --- greater modulation at lower energies --- 
but the difficulty of disentangling (in a limited dataset) two 
periodicities, considerable flickering, and the outburst decline,
 made further investigation unreliable. 

The X-ray data show the expected narrow, partial eclipse (not shown),
with a profile similar to that of the quiescent eclipse.
It is close to the time predicted by the ephemeris
of Hellier \&\ Sproats (1992; not including the sinusoidal term), 
being early by 40\,$\pm$\,10 s (0.007 in phase). Other than this, there
was no orbital modulation. 

\begin{figure} \vspace{6cm}      
\includegraphics{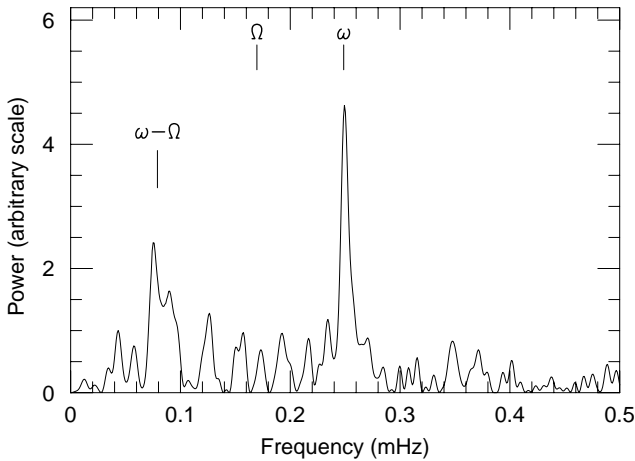}
\caption{The Fourier transform of the \xte\ data revealing modulations at the
spin ($\omega$) and beat (\beat) periods. }
\end{figure}

\subsection{CTIO photometry}
We obtained $R$-band photometry with the Cerro Tololo Inter-American 
Observatory 0.9-m telescope over 5 
nights of the decline and return to quiesence (Figs.~3 \&\ 5). 

The spin pulse is present throughout the dataset, but with different
amplitudes (the semi-amplitudes/mean, as far as can be told given the
flickering, are 12, 13, 24, 10 and 7 per cent on the five nights 
respectively). Reinsch \&\ Beuermann (1990) also report the pulsation
throughout their outburst dataset, with an amplitude comparable to that in
quiescence. 

It might appear from Fig.~5 that the pulse is late compared to
the predicted times of maxima (which use the quadratic ephemeris of
Hellier \&\ Sproats 1992) but the situation is more complex: Fig.~6
shows that X-ray maximum occurs where predicted (to within 0.05 in phase)
but that the optical pulse remains bright for \sqig 0.15 longer. This effect
has not been reported previously, but this is the first 
simultaneous optical/X-ray dataset. 

\begin{figure*} \vspace{10cm}      
\includegraphics{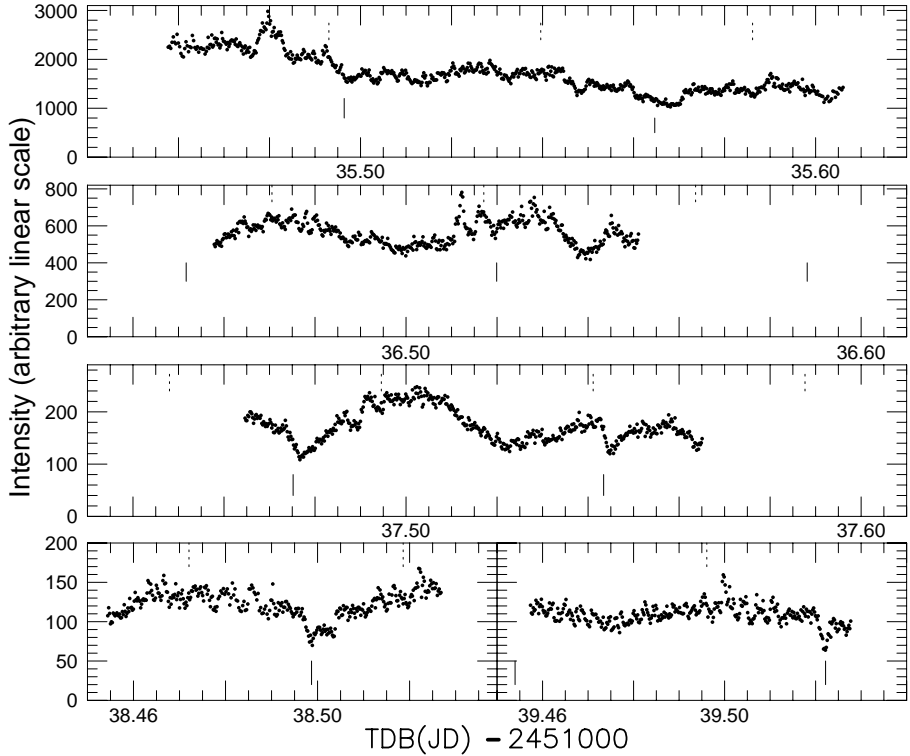}
\caption{The CTIO $R$-band photometry of EX~Hya on the decline from outburst
(note the differing $y$-axes). The lower tick marks show predicted eclipse 
times and the dotted ticks show times of spin maxima using the ephemerides
of Hellier \&\ Sproats (1992).}
\end{figure*}

\section{The quiescent eclipse}
Since interpreting the eclipse profiles during outburst will be crucial,
we'll first take a detour into the quiescent lightcurve.
Note, firstly, that the partial, flat-bottomed X-ray eclipse implies that the
secondary limb grazes the white dwarf, eclipsing the lower accreting pole 
but leaving the upper pole uneclipsed (Beuermann \&\ Osborne 1988;
Rosen \etal\ 1991; Mukai \etal\ 1998).

\begin{figure} \vspace{6.5cm}      
\includegraphics{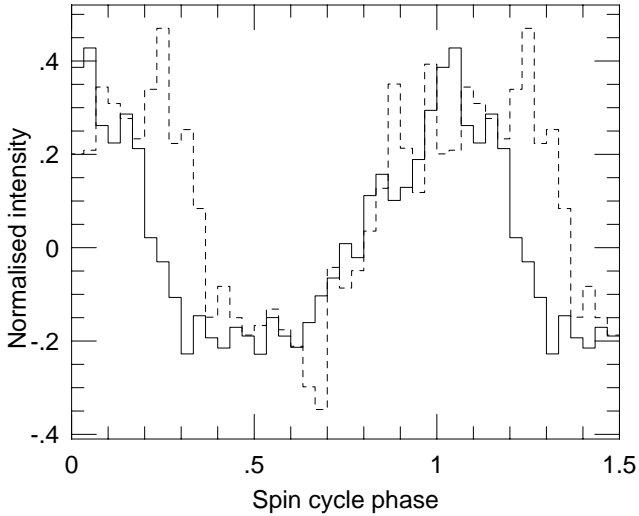}
\caption{The 2\up{nd}\ section of X-ray data (solid line) and the 
optical data (dashed) folded on the spin cycle.}
\end{figure}

To investigate the quiescent optical eclipse we have used the 45 h of
$B$-band photometry reported by Sterken \etal\ (1983) and Sterken \&\ Vogt
(1995). We first folded the data on the 67-min spin period (using 50 phase
bins) to obtain the mean pulse profile. We then removed the pulse by 
subtracting from each datapoint the value of the mean pulse profile at that phase. 
Then we folded the data on the orbital cycle, to obtain the curve displayed
in Fig.~7. The $\approx$\,30 per cent optical eclipse lasts for 3 mins and is
coincident with the X-ray eclipse.  Detailed studies (e.g.\ Siegel \etal\ 1989)
show that the eclipse centroid depends on spin phase and reveal that most of the
eclipsed light arises from the accretion curtain of material 
falling onto the lower pole of the white dwarf. 

Fig.~7 also shows an orbital hump extending between phases $\approx$\,0.6--0.15, 
and is presumably caused by the bright spot where the stream hits the accretion disc.
Similar features are seen in dwarf novae such as Z~Cha, where the hump extends
between phases 0.62--0.13 (Wood \etal\ 1986).

\begin{figure} \vspace{6.8cm}      
\includegraphics{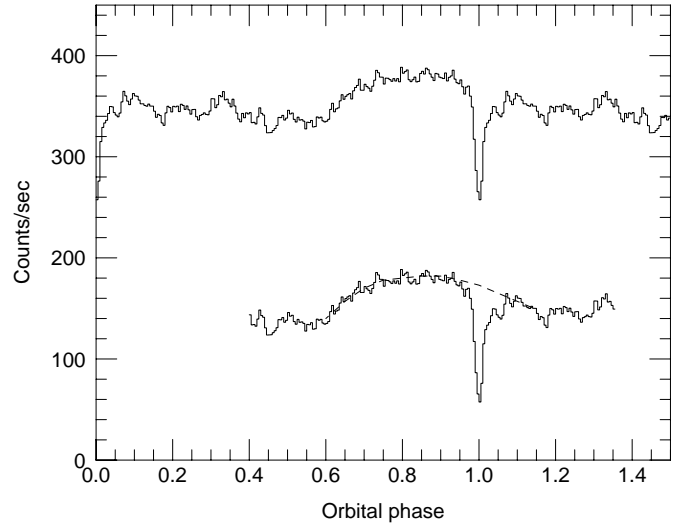}
\caption{The average quiescent orbital modulation from 45 h of $B$-band photometry.
The data are repeated, displaced downwards by 200 counts/sec, and with a
smoothed orbital hump added to guide the eye.}
\end{figure}

By analogy with Z~Cha, we would also expect to see a disc eclipse and a bright spot
eclipse. However, the disc eclipse involves only 23 per cent of the light in
Z~Cha (Wood \etal\ 1986), and in the grazing eclipse of EX~Hya the fraction
might be lower. It is possible that the EX~Hya light curve contains a disc
eclipse of $\approx$\,10 per cent depth, which starts at phase 0.95 as a steepening 
of the hump decline, and finishes at phase $\approx$\,0.05. It is also possible
that the `shoulder' to the eclipse, ending at
$\approx$\,0.07, involves the eclipse
of the bright spot. However, both interpretations are near the margins of the
data quality given the flickering. 

\begin{figure*} \vspace{6.5cm}      
\includegraphics{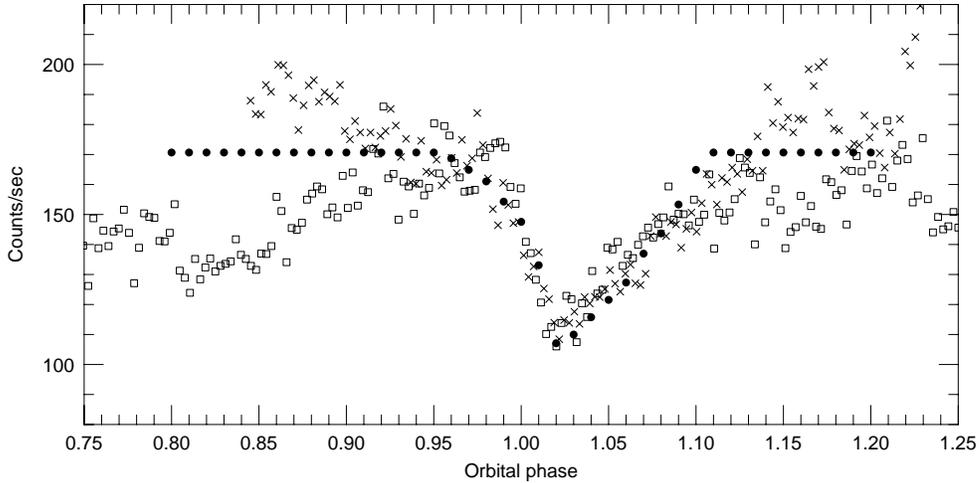}
\caption{The eclipse profiles from the third night of CTIO photometry
phased relative to the X-ray eclipse (crosses are from the first cycle,
squares from the second). The solid dots are a model eclipse of a bright
stream. Much of the out-of-eclipse variability is related to the spin
cycle.}
\end{figure*}

Since the evidence for a disc eclipse is marginal we can ask 
whether EX~Hya contains a disc at all, especially given the proposed models
of discless accretion in IPs (Wynn \&\ King 1995) and in EX~Hya in
particular (King \&\ Wynn 1999). The other evidence 
for a disc can be summarised as (see Hellier 1991 for a fuller account): (1) 
the dominance of the spin period, rather than the beat period, in quiescent 
X-ray lightcurves, which implies that the accreting material circularizes and 
loses knowledge of orbital phase; (2) a weak `rotational disturbance' seen in 
the emission lines (Paper~2); (3) an emission-line S-wave with the correct 
phase and velocity to arise from an impact at the edge of a disc (Paper~2); 
(4) the orbital hump and its being at the same phase as in Z~Cha (above); 
(5) material above the plane consistent with a splash where the stream hits the
disc edge, revealed by soft X-ray and EUV dips 
(C\'ordova, Mason \&\ Kahn 1985; Mauche 1999), 
and (6) the double-peaked lines seen in quiescence (Paper~2);

Note, though, that none of these secure the velocity field of the
disc, and so don't rule out a magnetically threaded structure
(e.g.\ King \&\ Wynn 1999) if it is able to mimic a disc in the
above respects. We leave the 
issue of whether the outburst was a disc instability, thus implying 
the presence of a disc, to the discussion. 

\section{The outburst eclipses}
In the last night of photometry (when \ex\ was back in quiescence) the 
observed eclipse was narrow, V-shaped, coincident with the X-ray 
eclipse, and similar to previous quiescent eclipses (Section~4).
On the penultimate night (JD 245\,1038) the eclipse egress had a `shoulder' 
lasting until phase 0.07. The night before, both eclipses had asymmetrical V shapes 
with minima at phase 
0.02 (relative to the X-ray mid-eclipse). Earlier still in the outburst the
eclipse is difficult to discern: some shallow dips may be eclipses, but 
there is not enough repeatability in consecutive cycles to distinguish them 
from flickering. Similarly, near the peak of the 1987 outburst Reinsch
\&\ Beuermann (1990) saw broad, shallow dips that may be eclipses, but
again there were not enough cycles to be sure. 

The lateness of the eclipses on the third night (JD 245\,1037) implies 
that they are probably eclipses of an accretion stream rather than a disc.
To test this we have computed the eclipse of a model stream, assuming it
to have a constant brightness along the freefall trajectory between the
initial impact with the disc and the point of its closest approach to the
white dwarf. The model parameters (based on those of Paper~2) are: 
$P_{\rm orb}$\,=\,5895 s, $M_{1}$\,=\,0.7 \msun, $M_{2}$\,=\,0.13 \msun,
$i$\,=\,79\deg\ and $R_{\rm disc}$\,=\,0.76 $R_{\rm L1}$. 

Fig.~8 shows that the model stream eclipse exhibits the same features
(V shape, minimum at phase 0.02, faster ingress, slower egress) as the
two eclipses observed on that night. The only free parameter is the model
normalization, where in order to match the data we have diluted the 
stream with uneclipsed light such that the stream is 43\%\ of the total 
[for comparison, in AM~Her stars the stream is commonly found to 
contribute 50--60\%\ of the total light (Harrop-Allin \etal\ 1999)].
Fig.~9 illustrates the geometry of the above model, showing the
system at phases 0.00 (white dwarf eclipse), 0.02 (stream-eclipse
minimum) and 0.07 (see below). 

On the fourth night (JD 245\,1038) the eclipse is of the white dwarf and 
its environs, with an additional shoulder lasting until phase 0.07
(unfortunately we only have one cycle that night so can't check the
feature's repeatability). The constant intensity during the shoulder
implies an eclipse of a pointlike source. If this source is 
located along the track of the accretion stream, the
start and end phases of the shoulder and the depth 
would be reproduced if it emitted 22\%\ of the system's light and was
0.29$a$ from the white dwarf (where $a$ is the stellar separation).

\begin{figure} \vspace{12.7cm}      
\includegraphics{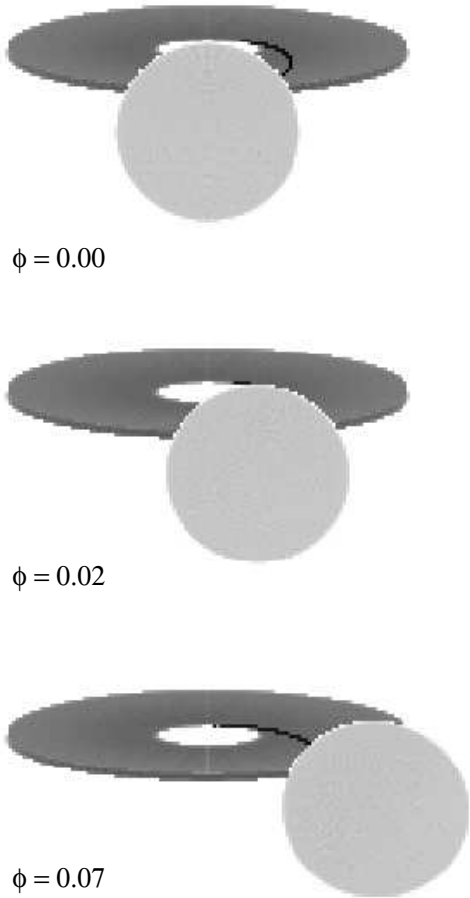}
\caption{An illustration of EX~Hya using the parameters discussed in the
text, shown at three phases during eclipse.}
\end{figure}

Is this distance (1.3\,\pten{10}\,cm) the radius of the magnetosphere? 
The (not particularly reliable) estimate of Paper~2 
is different, at 6\,\pten{9}\,cm.  However, we can check the plausibility
by estimating the field strength that would place the magnetosphere there.
Using the standard theory for a {\it disc\/} (e.g.\ Frank, King \&\ Raine
1992) we find that for an accretion rate of 10$^{16}$\,g\,s\minone\
the implied magnetic moment is 9\,\pten{32}\,G\,cm\up{3}, an order
of magnitude greater than other estimates (e.g.\ Paper~2; Warner 1996). 
One could argue that a stream would penetrate further in
than a disc, increasing the derived magnetic moment further, although
if the stream carried only a fraction of the accretion flow this would
reduce the estimate again.

Is the distance then the radius of the outer disc edge? 
Again, it is inconsistent with Paper~2, which found a value of  
2.3\,\pten{10}\,cm from the extent in phase of the rotational disturbance during 
the eclipse, and also from the separation of the double peaks in the 
emission lines, assuming these to give the Keplerian velocity at the disc edge.  
However, the extent of the rotational disturbance is hard to estimate, and the
assumption of Keplerian motion could well be wrong.  Further uncertainties are
that an enhanced stream might penetrate into the disc, and also that the 
disc size might change in outburst, enlarging due to the enhanced viscosity 
of a disc instability but shrinking due to the addition of low-angular-momentum
material from an enhanced stream.

Note, though, that the egress at phase 0.07 is consistent with a possible 
bright-spot egress at that phase in the quiescent lightcurve; thus, overall the 
most likely conclusion is that the feature is at the disc edge, and that 
previous estimates of the disc radius were too large.
 
\section{Discussion}
Several features of \ex 's outbursts are unlike those expected from 
a disc instability (Section~2). The most striking is the rarity of the 
outbursts. Over time, only 4\%\ of EX~Hya's accretion occurs during 
outburst (assuming, simplisticly, that the accretion rate scales as the optical flux). 
In contrast, the figure for a typical dwarf nova such as SS~Cyg is 90\%.
Another peculiarity is the range of interoutburst intervals,
from 8 d to $>$\,2 y, when there is no change in quiescent magnitude 
(and hence mass-transfer rate). Note also the decline times: the
2--3 d declines are typical of dwarf novae and are comparable with the
viscous timescale of a disc, but the 5-hr decline is not (even if allowance
is made for the lack of inner disc). 
Further, the emission line equivalent widths increase during outburst
(Paper~1); in dwarf novae (with the exception of IP~Peg) they decrease.

In contrast, the evidence for an enhanced mass-transfer stream is clear. 
High-velocity line wings from an overflowing stream hitting the
magnetosphere have been observed in both the 1987 and 1991 outbursts 
(Paper~1; Buckley \&\ Schwarzenberg-Czerny 1993). They were accompanied
by greatly enhanced line emission from the stream impact at the edge of 
the disc.  An X-ray beat period, caused by the stream connecting to the 
magnetic field and predicted in Paper~1, has now been seen (Section~3.1). 
Lastly, the eclipse profiles during late decline reveal a bright 
overflowing stream (Section~5).

The above suggests that EX~Hya's outbursts are mass transfer events,
rather than disc instabilities, but is not conclusive.
The disc-instability enthusiast could argue that the instabilities
are reduced in duration and frequency by the magnetic disruption of 
the inner disc (Angelini \&\ Verbunt 1989). If they are reduced to
minor perturbations on the disc, the irregularity of the outbursts
could follow. The enhanced mass transfer might then be a consequence
of a disc instability, triggered by enhanced irradiation of the secondary
star. This is easier in a magnetic system than in a dwarf nova since
radiation from the magnetic poles is less likely to be hidden by the disc, 
compared to radiation from a boundary layer.  There is indeed increased
line emission from the secondary in the 1987 outburst, in addition to the
increased line emission from the stream/bright-spot (Paper~1). 

In principle the optical eclipse profiles early in the outburst should 
tell us whether the disc has gone into a high state, however the difficulty
of judging which features are real, given the flickering and limited data,
precludes a firm conclusion. The first two nights of our optical dataset,
and also the dataset early in an outburst by Reinsch
\&\ Beuermann (1990) are compatible with broad dips, \sqig 20 per cent 
deep, at the expected eclipse times. If a disc were the only light 
source, the grazing eclipse would produce dips of \sqig 30 per cent
depth, so the observed depth is consistent with some dilution by light
from the magnetosphere (which must be present given the spin pulse). 
The broad dips might also be centered slightly late, compared to the
expected eclipse time (Reinsch \&\ Beuermann 1990, and also 
upper panel of our Fig.~5), which would indicate a contribution from an 
enhanced stream. 

The absence, early in the outburst, of the narrow
eclipses seen in quiescence is puzzling. It indicates either that the
white dwarf and its accretion curtains are relatively faint (but why then
do we still see a spin pulse?) or that in outburst we see predominantly
the upper (uneclipsed) accretion curtain (but why is this?). 

In summary, there is clearly enhanced mass transfer, but we cannot be sure
whether or not this is triggered by a disc-instability outburst.

\section{Conclusions}
(1) \ex 's outbursts are unlike those of any other dwarf novae. 
However, it is possible that they are characteristic of disc instabilities in
a magnetically truncated disc rather than the result of a different process. 

(2) There is clearly enhanced mass transfer in outburst. The evidence
includes an enhanced stream/disc impact, eclipse profiles resulting from
a bright stream overflowing the disc, line emission from where the overflowing
stream hits the magnetosphere, and an X-ray periodicity at the beat period, 
indicating coupling of the overflowing stream to the magnetosphere.

(3) It is unclear whether the enhanced mass transfer is triggered by a disc
instability. The eclipse profiles early in outburst are consistent with a 
brightened disc, but there isn't enough repeatability over different cycles
to distinguish them from flickering with certainty.

(4) After reviewing the evidence we conclude that EX~Hya does possess an
accretion disc, or a circulating structure with very similar
characteristics. 
The optical orbital modulation in quiescence is similar to that of 
non-magnetic dwarf novae, including an orbital hump and marginal evidence 
for eclipses of the disc and bright spot.

(5) We find evidence that previous estimates for the disc size are too large,
prefering instead a disc radius of 0.29 of the stellar separation. 

(6) In possessing an overflowing stream giving rise to distorted emission line
wings and distorted eclipse profiles, \ex\ in outburst shows some of the
characteristics of SW~Sex stars (e.g.\ Hellier 1999).

\section*{Acknowledgments}
We thank Darragh O'Donoghue for suggesting the analysis of Section~4,
Chris Sterken for kindly sending us the data, and Janet Wood for helping
us to interpret it.  We thank the \xte\ team for their rapid response
to our TOO request. Further, we thank Klaus Beuermann for a helpful
referee's report. 
The Cerro Tololo Inter-American Observatory, National
Optical Astronomy Observatories, is operated by the Association of
Universities for Research in Astronomy, Inc.\ (AURA) under cooperative
agreement with the National Science Foundation.

{}

\begin{thebibliography}{}
\bibitem[]{}van Amerongen S., van Paradijs J., 1989, A\&A, 219, 195
\bibitem[]{}Angelini L., Verbunt F., 1989, MNRAS, 238, 697
\bibitem[]{}Beuermann K., Osborne J.\,P., 1988, A\&A, 189, 128
\bibitem[]{}Bond I.\,A., Freeth R.\,V., Marino B.\,F., Walker W.\,S.\,G.
    1987, IBVS No.~3037
\bibitem[]{}Bradt H.\,V., Rothschild R.\,E., Swank J.\,H., 1993, A\&AS, 97, 355
\bibitem[]{}Buckley D.\,A.\,H., Schwarzenberg-Czerny, A., 1993, 
        Annals of the Israeli Physical Society, 10, 278
\bibitem[]{}C\'ordova F.\,A., Mason K.\,O., Kahn S.\,M., 1985, MNRAS,
            212, 447
\bibitem[]{}Frank J., King A. R., Raine D. J., 1992, Accretion power in 
            astrophysics, Cambridge University Press, Cambridge.

\bibitem[]{}Harrop-Allin M.\,K., Cropper M., Hakala P.\,J., Hellier C.,
              Ramseyer T., 1999, MNRAS, 308, 807
\bibitem[]{}Hellier C., 1991. MNRAS, 251, 693
\bibitem[]{}Hellier C. 1998, Adv.\ Space Res., 22(7), 973
\bibitem[]{}Hellier C., 1999, in Charles P.\,A., King A., O'Donoghue D., eds,
           ``Proceedings of the Warner Symposium on cataclysmic variables'',  
           New Astr.\ Rev., in press 
\bibitem[]{}Hellier C., Buckley D.\,A.\,H., 1993, MNRAS, 265, 766
\bibitem[]{}Hellier C., Mason K.\,O., Rosen S.\,R., C\'ordova F.\,A, 1987,
     MNRAS, 228, 463
\bibitem[]{}Hellier C., Mason K.\,O., Smale A.\,P., Corbet R.\,H.\,D.,
    O'Donoghue D., Barrett P.\,E., Warner B., 1989, MNRAS, 238, 1107
\bibitem[]{}Hellier C., Mukai K., Beardmore A.\,P., 1997, MNRAS, 292, 397
\bibitem[]{}Hellier C., Sproats L.\,N., 1992, IBVS, No.~3724
\bibitem[]{}Kato T., vsnet-recent 15156, 
http://www.kusastro.kyoto-u.ac.jp/vsnet/Mail/recent15000/msg00156.html
\bibitem[]{}Kim S.-W., Wheeler J.\,C., Mineshige S., 1992, ApJ, 384, 269
\bibitem[]{}King A.\,R., Wynn G.\,A., 1999, MNRAS, 310, 203
\bibitem[]{}Mauche C.\,W., 1999, ApJ, 520, 822
\bibitem[]{}Mukai K., Ishida M., Osborne J., Rosen S., Stavroyiannopoulos D.,
   1998, in Howell S., Kuulkers E., Woodward C., eds, ``Wild stars in the old
west'',  ASP Conf.\ Ser., 137, 554 
\bibitem[]{}Patterson J., Schwartz D.\,A., Pye J.\,P., Blair W.\,P.,
            Williams G.\,A., Caillault J.-P., 1992, ApJ, 392, 233
\bibitem[]{}Reinsch K., Beuermann K., 1990., A\&A, 240, 360
\bibitem[]{}Rosen S.\,R., Mason K.\,O., Mukai K., Williams O.\,R., 1991,
    MNRAS, 249, 417
\bibitem[]{}Schwarz, H., van Amerongen, S., Heemskerk, M.\,H.\,M.,
            van Paradijs, J., 1988, A\&A, 202, L16
\bibitem[]{}Siegel N., Reinsch K., Beuermann K., van der Woerd H., Wolff E.,
    1989, A\&A, 225 97
\bibitem[]{}Sterken C., Vogt N., 1995, J.\ Astr.\ Data, 1, 1
\bibitem[]{}Sterken C., Vogt N., Freeth R., Kennedy H.\,D., Marino B.\,F.,
    Page A.\,A., Walker W.\,S.\,G., 1983, A\&A, 118,  325
\bibitem[]{}Szkody P., Mateo M., 1984. ApJ, 280, 729.
\bibitem[]{}Tovmassian G.H. \etal\ 1998, A\&A, 335, 227
\bibitem[]{}Warner B., 1996, Ap\&SS, 241, 263
\bibitem[]{}Warren J.\,K., Vallerga J.\,V., Mauche C.\,W., Mukai K.,
            Siegmund O.\,H.\,W., 1993, ApJ, 414, L69.
\bibitem[]{}Wood J.\,H., Horne K., Berriman G., Wade R., O'Donoghue D., 
              Warner B., 1986, MNRAS, 219, 629
\bibitem[]{}Wynn G.\,A., King A.\,R., 1995, MNRAS, 275, 9
\end{thebibliography}
\end{document}